\documentclass[%
reprint,
superscriptaddress,
groupedaddress,
nofootinbib,
amsmath,amssymb,
aps,
prstab,
floatfix,
]{revtex4-2}

\usepackage{dblfloatfix}
\usepackage{graphicx}
\usepackage{dcolumn}
\usepackage{bm}
\usepackage{siunitx}
\usepackage{tablefootnote}
\usepackage{placeins}
\usepackage{xcolor}
\usepackage{appendix}
\usepackage{multirow}
\usepackage{hyperref}

\begin{document}

\preprint{APS/123-QED}

\title{Space Charge and 
Future Light Sources}

\author{S.~A.~Antipov\textsuperscript{1}\thanks{Sergey.Antipov@desy.de}
}

\author{V.~Gubaidulin\textsuperscript{2}}

\author{I.~Agapov\textsuperscript{1}
}
\author{E.~C.~{Cort{\'e}s Garc{\'i}a}\textsuperscript{1}
}

\author{A.~Gamelin\textsuperscript{2}}

\affiliation{\textsuperscript{1}Deutsches Elektronen-Synchrotron DESY, Notkestr. 85, 22607 Hamburg, Germany}

\affiliation{\textsuperscript{2}Synchrotron SOLEIL, Saint-Aubin 91190, France}

\date{\today}

\begin{abstract}

It is a truth universally acknowledged, that space charge effects in ultrarelativistic electron storage rings are irrelevant due to the steep inverse dependence of their strength on the Lorentz factor. Yet, with the push towards the diffraction limit, the state-of-the-art light sources are approaching the point where their emittance becomes so small that the space charge force can no longer be ignored. In this paper, we demonstrate how space charge effects affect the injection dynamics, dynamical aperture, and collective beam stability on the example of 4th generation light sources PETRA~IV and SOLEIL~II.


\end{abstract}

\maketitle

\section{Introduction}

Accelerator lattices are typically being designed with an aim at optimizing single particle dynamics. But when a charged particle orbits the accelerator it sees the neighboring particles and interacts with them, the phenomenon known as the space charge (SC) effect. This Coulomb interaction, nonlinear in nature, may lead to many unwanted consequences. 

The nonlinear space charge force alters single particle dynamics in the ring, shifting the frequencies of transverse oscillations around the reference orbit (i.e. the incoherent tunes). It may lead to resonance excitation, emittance growth, halo formation, and losses~\cite{Weng:1986ik, Bartosik:2020ger, Oeftiger:2021klb}. Space charge also affects the collective dynamics in many ways. It has an effect on the frequencies of head-tail modes and can suppress transverse mode coupling instability~\cite{ABS_Model}.
At the same time, since it shifts the incoherent tunes, it also has an impact on Landau damping and suppression of head-tail instabilities~\cite{kornilov_head-tail_2010,Kornilov:2020bwt,balbekov_transverse_2009,burov_head-tail_2009, burov_erratum_2009}. 
Lastly, sufficiently strong SC can give rise to convective instabilities within high-intensity bunches~\cite{Burov_22}.

The SC effects are the most prominent in the frontier high-intensity hadron machines, e.g. SPS (CERN), SIS100 (FAIR/GSI) and their injectors. Lepton rings, on the other hand, benefit from the much smaller mass of the particles as the SC interaction decreases with the Lorentz factor, which is typically orders of magnitude larger than in hadron rings. Yet, it has been noted that SC effects might be important in the damping rings of a linear lepton collider where a combination of relatively low beam energy and small emittance makes SC interaction significant. SC has been found problematic, affecting coupling and low-order resonances in both ILC~\cite{Decking:2000ka, Xiao:2007lff} and CLIC~\cite{Rumolo:2008zzb, ZampetakisPRAB2024} designs. 

\begin{figure}
    \centering
    \includegraphics[width = \linewidth]{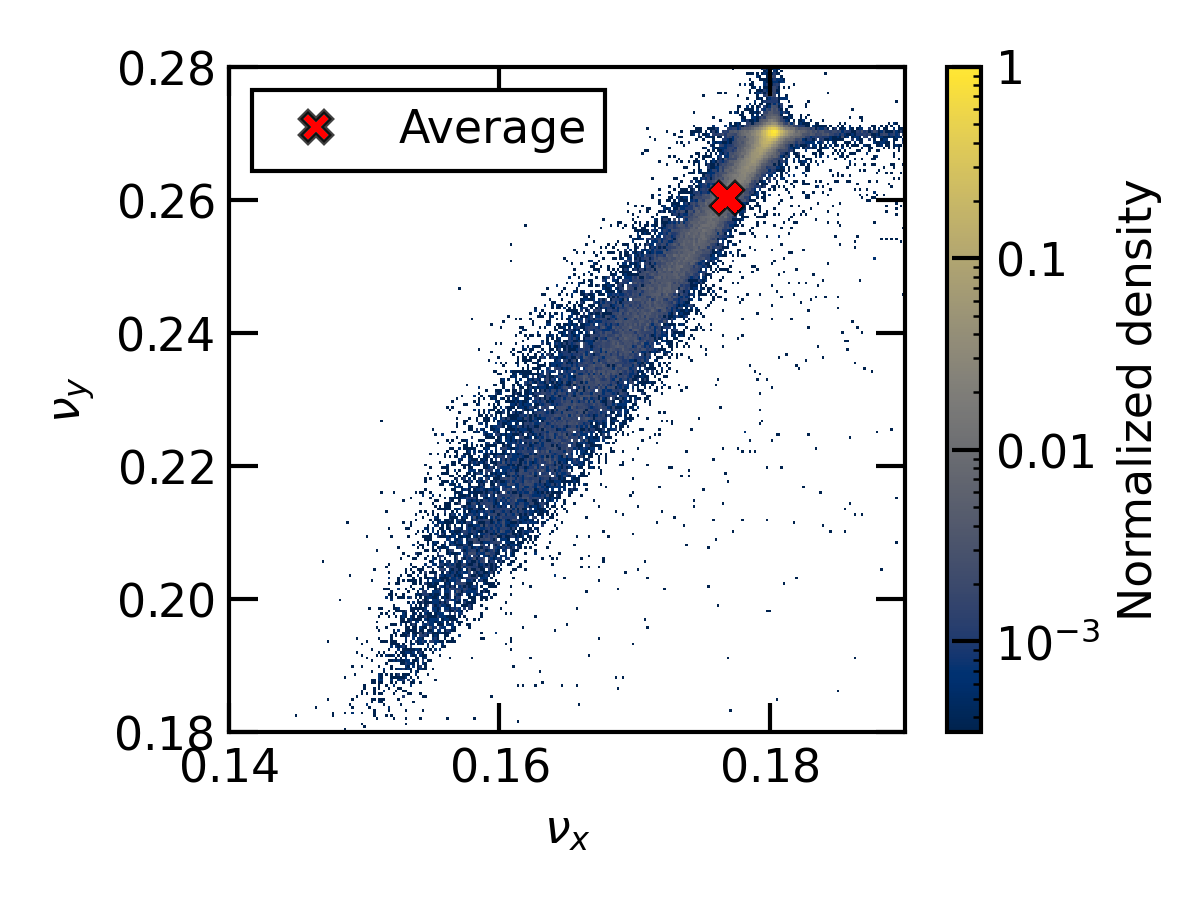}
    \caption{Exemplary tune footprint with space charge of a 10~nC bunch in PETRA~IV. Incoherent particle tunes spread far away from the nominal working point of (0.18, 0.27). Numerical simulation in \texttt{XSuite}~\cite{XSuite}.}
    \label{fig:tune-footprint-PETRA-IV}
\end{figure}

Contemporary synchrotron light sources are few-hundred-meter- to kilometer-long electron storage rings, operating at energies around a few GeV. By design, they feature relatively large circumferences, necessary for achieving the smallest beam emittance and, thus, the largest photon brilliance. With the latest advances in multibend achromat focusing optics~\cite{Einfeld:xe5006, Raimondi2023} the state-of-the-art 4th generation light sources reach geometric emittances in the $10-100$~pm range. At this level, the particle density within the bunch becomes so large that the SC effects might become important~(Fig.~\ref{fig:tune-footprint-PETRA-IV}). Nagaoka and Bane~\cite{nagaoka_collective_2014} predicted that the incoherent optics distortion produced by SC may have negative effects on machine performance, such as emittance increase, reduction of injection efficiency and beam lifetime. Nevertheless, the topic of SC is frequently overlooked at the design stage of future light sources. 

In this paper, we demonstrate that SC does indeed become significant for the 4th generation synchrotron light sources, both for single particle and for the collective dynamics. It can have an effect on various aspects of machine performance, from beam emittance to the intensity limit. We analyse its impact in detail on the examples of two 4th generation light sources that cover the energy range of most future machines: a high energy 6~GeV PETRA~IV light source and a medium energy 2.75~GeV SOLEIL~II.



\newpage

\section{Space charge effects}

Let us begin by introducing our two study cases. Both machines feature multibend achromat lattices~\cite{Einfeld:xe5006, Raimondi2023} with tight focusing and beam sizes reaching as low as $\sigma \lesssim 10~\mu$m. The 6 GeV PETRA~IV machine utilizes a hybrid H6BA lattice~\cite{agapov2024beamdynamicsperformanceproposed} with a large dynamic aperture and beam lifetime. It will be equipped with fast kickers for off-axis top-up injection. SOLEIL~II is a more compact machine operating at a \SI{2.75}{\giga\eV} energy, with its higher order achromat lattice combining 4BA and 7BA cells~\cite{nadji_upgrade_2023, loulergue_tdr_2023}. It is designed to operate with a multipole injection kicker (MIK) for transparent injection~\cite{alex:ipac23-thpa175}. Both machines have a double rf system comprising of the main rf and a higher harmonic rf to stretch the bunches and increase the beam lifetime. PETRA~IV will utilize active normal conducting  500~MHz main and 1.5~GHz 3rd harmonic cavities, while SOLEIL~II will use 352.20~MHz main cavities and 1.4~GHz passive normal conducting 4th harmonic cavities. The machines envisage delivering to users both a high-beam-current, low-charge Brightness mode as well as a high-charge Timing mode with a smaller number of bunches. Table~\ref{tab:P4_params} lists the key parameters of both machines.

\subsection{Space charge tune spread}

Consider a beam of short relativistic electron bunches (charge $e$, mass $m_e$) orbiting the storage ring of a light source. The Coulomb interaction between the particles will change their natural frequencies of oscillation around the reference orbit, causing so-called space charge tune shift. In the linear model, the maximal tune shift of a Gaussian bunch can be estimated as
\begin{equation}
    \Delta \nu^{SC}_{x,y} = - \frac{N r_e C}{(2\pi)^{3/2}\gamma^3\sigma_z}
    \biggl \langle 
    \frac{\beta_{x,y}}{\sigma_{x,y}(\sigma_x+\sigma_y)} 
    \biggr \rangle,
    \label{eq:SC_tune_shift}
\end{equation}
where $N$ is the number of particles per bunch, $\sigma_{x,y,z}$ are the rms horizontal, vertical and longitudinal beam sizes, $\gamma$ is the Lorentz factor, $\beta_{x,y}$ are the transverse Twiss optics functions, and $\langle ... \rangle$ denotes averaging over the ring circumference $C$. $r_e = e^2 / m_ec^2 \approx 2.8\times10^{-13}$~cm (in cgs units) stands the classical electron radius and $c$ is the speed of light. In a typical scenario of a machine operating with flat beams, $\sigma_y \ll \sigma_x$, the largest tune shift occurs in the vertical, $y$-plane.

While the $1/\gamma^3$ factor can be as small as $10^{-12}$ in GeV-range light source, it is compensated by the second term in Eq.~(\ref{eq:SC_tune_shift}). The small beam size may result in a SC tune shift of the order of or larger than the synchrotron tune.
For example, in PETRA~IV if operated without the 3rd harmonic cavities, the SC tune shifts reach $10^{-2}$ level for the low-charge Brightness mode and nearly 0.1 for the high-charge Timing mode (Table~\ref{tab:SC_tune_shifts}). This is significant and might affect the single-particle nonlinear dynamics in the ring. A measure of SC impact on the collective instabilities is the SC parameter, i.e. the ratio of its tune shift to the synchrotron tune $|\Delta \nu_y^{SC}| / \nu_s$. At $|\Delta \nu_y^{SC}| / \nu_s \sim 5-20$, depending on the bunch charge, the SC is expected to play a role in the collective dynamics as well. With the bunch lengthening 3rd harmonic cavities the SC tune shift decreases to $1-4\times 10^{-2}$, still a significant figure.

Note that the case of PETRA~IV is somewhat special as, inheriting its ring tunnel from the high energy physics program, PETRA features a relatively large circumference for its energy, resulting in a large SC tune shift. This would also be the case for other ex-collider machines, such as a potential ring in the PEP tunnel at Stanford~\cite{BEI2010518, RAIMONDI2024169137}. `More conventional' light sources normally have a circumference between 300 and 500~m for $\sim 3$~GeV machines and about 1~km for 6~GeV machines. For instance, a 4th generation 2.75~GeV light source SOLEIL~II will have a 354~m circumference (Table~\ref{tab:P4_params}). Its bright beams will have $|\Delta \nu_y^{SC}| / \nu_s \sim 2-25$ without a harmonic cavity.

In the present-day 3rd generation light sources, the SC tune shifts are significantly smaller thanks to the much greater transverse emittance.
For example, PETRA~III has $\epsilon_x \approx 1$~nm with the highest bunch charge of 19~nC in a 40-bunch timing mode~\cite{Balewski:392205}. Under these conditions, the SC tune shift is $\sim -6\times 10^{-3}$, while $\nu_s = 0.049$ and the SC parameter $|\Delta \nu_y^\text{SC}| / \nu_s \approx 0.1$, a rather small quantity to have any meaningful effect on the beam.
In SOLEIL~\cite{filhol_overview_2006}, $\epsilon_x \approx 3.9$~nm and $\Delta \nu_y^\text{SC} \sim \num{-1.4e-3}$ for nominal charge of \SI{1.4}{\nano\coulomb} and $\Delta \nu_y^\text{SC} \sim \num{-5e-3}$ for highest possible charge of $\sim \SI{25}{\nano\coulomb}$, while $\nu_s=\num{4.8e-3}$.
The SC parameter in the case of SOLEIL is $|\Delta\nu_y^\text{SC}|/\nu_s \lesssim 0.1 - 1$.
In this case, there should not be a significant effect on the beam dynamics.

\begin{table}
    \caption{PETRA~IV and SOLEIL~II machine parameters and their main baseline operation modes: Brightness and Timing (in parenthesis).}
    \centering
    \begin{tabular}{lrcc}
        \hline \hline
        Parameter & Symbol & PETRA~IV & SOLEIL~II\\
        \hline
         Energy (GeV) & $E$  &  6.0 & 2.75\\
         Circumference (m) &$C$  & 2304 & 354\\ 
         Momentum compaction & $\alpha_C$ & $3.3\times 10^{-5}$ & \num{1.07e-04}\\
         Betatron tunes & $\nu_{x,y}$ & 135.18, 86.27 & 54.2, 18.3\\
         Operational chromaticities & $\xi_{x,y}$ & 6, 6 & 1.6, 1.6\\
         Main rf frequency (MHz) & $f_{\text{rf}}$ & 500.0 & 352.20 \\
         Synchrotron tune & $\nu_s$\footnote{without the harmonic rf system} & 0.005 & 0.002\\
         Bunch charge (\unit{\nano\coulomb}) & $Q_b$ & 1 (8) & 1.4 (7.4)\\
         Rms bunch length (\unit{\pico\second}) & $\sigma_t$ & 40 (65) & 42 (50)\\
         Emittance (\unit{\pico \meter \radian}) & $\epsilon_0$ & 20 & 83\\
         SR damping times (ms)& $\tau_{x,y,s}$ & 18, 22, 13 & 8, 14, 12 \\  
         \hline
    \end{tabular}
    \label{tab:P4_params}
\end{table}
\newpage
\subsection{Longitudinal space charge}

It is worth mentioning that while the transverse SC might be significant thanks to the small beam size, the same is not the case for the longitudinal SC. Indeed, the synchrotron SC tune shift scales with $\ln(b / \sigma_y)$, where $b$ is the radius of the vacuum chamber~\cite{OeftigerPhDThesis}. This logarithm does not exceed $\sim 10$, and the tune shift is 
\begin{equation}
    \Delta \nu_s^\text{SC} \sim - \frac{N \alpha_C r_e C^2}{8\pi^2\gamma^3 \nu_s \sigma_z^3}.
\end{equation}
For the beam parameters in Table~\ref{tab:P4_params}, $\Delta \nu_s^\text{SC}\lesssim 10^{-7}$ which is a negligibly small quantity.

\subsection{Space charge-driven resonances}
Space charge force, non-linear in nature, produces a reach variety of resonance conditions in a machine~\cite{Hofmann:2017isl}. In particular, because the SC force depends on the longitudinal position of a particle within the bunch, a particle performing synchrotron oscillations will, therefore, experience a tune modulation with twice the synchrotron tune $\nu_s$. For a flat beam where the vertical SC force dominates this leads to a resonance condition~\cite{Decking:2000ka}
\begin{equation}
    n\nu_x + m(\nu_y \pm 2\nu_s) = l,
    \label{eq:SC_resonance}
\end{equation}
where $l,m,n \in \mathbb{Z}$.
In principle, the presence of SC-specific resonance lines might affect the boundary of the chaotic region in the phase space, limiting the dynamic aperture of the machine and leading to injection losses. However, the SC effect is only experienced by particles that are in the core of the stored bunch. Particles still exerting ample betatron oscillations (e.g. after a top-up injection) will not be affected by it. 

\subsection{Mitigation of transverse mode coupling instability}
Another important effect of SC is its shift of frequencies of coherent head-tail modes, which manifests itself in the mitigation of transverse mode coupling instability (TMCI) of modes \num{0} and \num{-1}~\cite{ABS_Model}. The TMCI limits the bunch charge at chromaticity 0 (here, and later, we denote \textit{unnormalized} chromaticity as $\xi$). It is characterized by a fast rise time of the order of a synchrotron period, which is typically much faster than the synchrotron radiation (SR) damping time. A normal resistive feedback is also inefficient at suppressing the instability and, in fact, may have a destabilizing effect~\cite{elias:fb}. While TMCI can be mitigated by a simple increase of $\xi$, there may be other considerations in favour of working at a low chromaticity, such as a smaller sextupole strength, which might result in a larger DA, better injection efficiency, and lower power consumption.\footnote{While in simulations the DA of PETRA~IV does not change significantly with chromaticity, this might not be the case in a real machine due to magnet errors and imperfections. Experience on a 4th generation ESRF EBS light source, which utilizes a similar HMBA lattice, suggests the presence of an optimal sextupole setting that maximizes beam lifetime~\cite{Liuzzo:2023sud}.} The TMCI threshold for a zero chromaticity lattice is approximately~\cite{Handbook:SB_Instab}
\begin{align}
Q_\text{th} = 8 \pi^{3/2} \frac{E}{e}\frac{\nu_s \sigma_z}{C} \min \left(\frac{\nu_{x,y}}{Z_{x,y}}\right),\label{eq:TMCI}
\end{align}
where $\nu_s$ is the synchrotron tune, $Z_{x,y}$ is the effective transverse impedance, and $E$ is the beam energy. Typically, in a light source it is the $y$-plane that poses the limitation due the larger beam coupling impedance because of the narrow undulator chambers. The formula is valid for a single harmonic rf system, assuming the bunch lengthening from longitudinal impedance is small.

\subsection{Landau damping by space charge}
The large nonlinear betatron tune spread generated by SC gives rise to Landau damping of collective head-tail modes in the electron bunches. From the results obtained for hadron synchrotrons analytically \cite{balbekov_transverse_2009, burov_head-tail_2009, burov_erratum_2009} and in simulations \cite{kornilov_head-tail_2010, macridin_simulation_2015} one might expect a strong suppression of head-tail instabilities by SC in the region $|\Delta\nu^\text{SC}_y|/\nu_s\in [2, 20]$.

A few notable differences exist between the high-intensity hadron machines and low-emittance electron synchrotrons though.
Firstly, the low-emittance ring design dictates strong focusing, resulting in smaller vacuum chamber dimensions and stronger impedance effects.
This leads to impedance-induced coherent tune shifts of the order of $\nu_s$.
Thus, individual head-tail modes cannot be treated independently. 
Secondly, the transverse wakefield force cannot be assumed to be much smaller than the space charge force, as is the case in~\cite{balbekov_transverse_2009, burov_head-tail_2009, burov_erratum_2009,kornilov_head-tail_2010, macridin_simulation_2015}. Consequently, the impedance-induced tune shifts cannot be assumed to be much smaller than those created by SC. 
Lastly, the electron storage rings are more susceptible to the high-frequency impedance components due to their shorter bunch length: \num{10} -- \SI{100}{\pico\second} vs \num{1} -- \SI{1000}{\nano\second} in hadron machines.

\begin{table}[t]\caption{Estimated SC tune shifts in PETRA~IV (PIV) and SOLEIL~II (SII)}
    \begin{tabular}{l c c c c c}
    \hline\hline
    \multirow{2}{*}{} & $Q_b$ & \multicolumn{2}{c}{w/ harm cav} & \multicolumn{2}{c}{w/o harm cav}\\
    & (nC) & $\Delta \nu_x$ & $\Delta \nu_y$ & $\Delta \nu_x$ & $\Delta \nu_y$ \\
    \hline
    
    \multirow{2}{*}{\rotatebox[origin=c]{90}{PIV}}&
    1 & \num{-2.8e-3}& \num{-1.0e-2} & \num{-7.6e-3}& \num{-2.7e-2} \\
    & 8 & \num{-1.2e-2}& \num{-4.1e-2} & \num{-2.2e-2}& \num{-7.7e-2}\\
    \hline

    \multirow{2}{*}{\rotatebox[origin=c]{90}{SII}} &
    1.4 & \num{-3.1e-3} & \num{-6.2e-3} & \num{-8.8e-3}  & \num{-1.8e-2} \\
    & 7.4 & \num{-1.4e-2} & \num{-2.9e-2} & \num{-2.7e-2} & \num{-5.4e-2} \\
\hline
    \end{tabular}
    \label{tab:SC_tune_shifts}
\end{table}

\section{PETRA~IV case}\label{sec:P4_case}

As the first example of SC affecting the beam dynamics, let us consider the case of the 6 GeV PETRA~IV machine in more detail. To study the effect of SC on PETRA~IV, simulation campaigns were carried out with \texttt{ELEGANT}~\cite{Borland:2000gvh}. We investigated the impact on the dynamic aperture, the dynamics of the off-axis top-up injection, and on the single-bunch charge accumulation limit. 

\subsection{Simulation setup}
Our simulation setup included 6D element-by-element tracking, effects of synchrotron radiation, quantum excitation, longitudinal and transverse dipolar and quadrupolar impedance. 
The impedance model accounts for all major sources of wakefields in the ring the vacuum chambers, rf cavities, injection and feedback kickers, BPMs, current monitors, tapers, bellows, and radiation absorbers. The resistive wall contributions of components were computed with \texttt{IW2D}, including the planned NEG coating of vacuum chambers (resistivity of 2~$\mu\Omega$m, according to rf measurements of coated tube samples). The geometric part of the impedance has been computed using \texttt{CST Particle Studio} and \texttt{GdfidL} codes. 

Misalignment and gradient field errors were introduced to generate an rms $\beta$-beat of $\sim \SI{3}{\%}$. 
 The coupled bunch dynamics was outside of the scope of this first SC study due to the computational complexity of the problem (1920 bunches in the Brightness mode). The intrabeam scattering (IBS) was taken into account indirectly by initializing the beam with charge-dependent distribution parameters, obtained from a separate study of IBS.
 This approach is justified when focusing on the onset of collective instabilities and simulating the time scales of about an SR damping time. At these time scales, the distribution should not have modified significantly due to the absence of IBS from our tracking, thanks to the high beam energy, which reduced the IBS growth rates.\footnote{The overall impact of IBS on the equilibrium horizontal emittance is below 5~pm in the Brightness mode with harmonic cavities~\cite{Bartolini:2022bor}.}
The direct SC element in \texttt{ELEGANT} follows a quasi-frozen model and only includes transversal non-linear effects. The model utilizes Basetti-Erskine approximation~\cite{bassetti_closed_1980} to compute the electric field of an ultrarelativistic Gaussian bunch and updates the kick strength according to the local beam parameters~\cite{Xiao:2007lff}. SC kicks were added at every quadrupole to simulate the effect at each turn, while also sampling the evolution of the beam size around the ring. This is a sufficient number of kicks in order to achieve numerical convergence in terms of equilibrium beam emittances (App.~\ref{App:B}).

In addition, \texttt{XSuite}~\cite{XSuite} was used to run tracking simulations for an independent benchmark.
The tracking configuration retains all of the effects from \texttt{ELEGANT}, except for misalignment and gradient field errors. 
Three types of SC kicks in \texttt{XSuite} are available: particle-in-cell (PIC), frozen, and quasi-frozen.
SC kicks with the PIC model were inserted into the lattice at equal intervals to support the findings from the \texttt{ELEGANT}.
Benchmarks are included in App.~\ref{App:C}.
If not otherwise stated, SC kicks in \texttt{XSuite} were introduced with the quasi-frozen model.

\subsection{Dynamics of top-up injection}

\begin{figure}
    \centering
    \includegraphics[width = 0.99\linewidth]{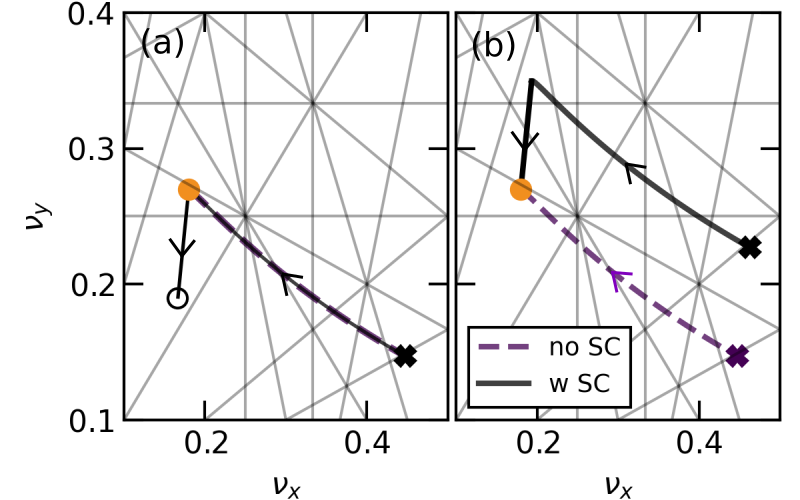}
    \caption{Evolution of particle tune during the off-axis injection process with and without taking the SC force into account. (a) -- nominal tunes; the SC shifts the average tune of the stored bunch (black circle) from its set point (orange dot). (b) -- assuming a tune correction applied to bring the average tune with SC to the set point value (orange dot); the tune correction affects the tune of the bunch injected off-axis (cross). The curves represent the excursion of off-axis injection of a flat beam, the crosses represent the extension of a particle at 3~rms from the injected bunch centroid. Bunch charge 20~nC, bunch lengthening by longitudinal wakefields and 3rd harmonic rf included.
    }
    \label{fig:tune-injection-dynamics}
\end{figure}

In the case of PETRA IV, pursuing an off-axis injection scheme, the dynamics is, in first instance, dictated by the strong amplitude detuning imposed by the HMBA lattice. Particles in the injected bunch possess a distinct tune value, far from the nominal. 
While the tune depression described by Eq.~(\ref{eq:SC_tune_shift}) is incoherent, it effectively spreads the tune distribution and shifts its average. This implies that under strong SC influence, the injected bunch can find itself at a different working point. 
To account for this effect, one could naively shift the nominal working point. But while a tune correction (e.g. with a tune feedback system) could restore the
average tunes of the machine, significantly detuned particles will still remain. Moreover, the same correction would affect the tunes of the injected bunch during the off-axis injection.
The tune of the injected bunch is sketched in Fig.~\ref{fig:tune-injection-dynamics}. SC affects the the trajectory of the injected bunch on the tune diagram as it damps down to the reference orbit. When applying tune correction, the injected bunch would trace a different path in depending on the charge of the stored bunch. Also, considering the stop band width excited by misalignments and gradient field errors, displacing the working point might reduce the injection efficiency.
This can be seen in Fig.~ ~\ref{fig:tune-injection-dynamics}b, where a tune correction intended to bring the tunes of the stored bunch back to its nominal working point displaces the tunes of the injected bunch towards a half-integer resonance\footnote{In reality, the width of the resonance stop band might differ depending on the actual errors and misalignments of a real machine.}.
This way the presence of charge-dependent tune shift may pose constraints on the tune correction and injection amplitude.

\subsection{Dynamic aperture}

\begin{figure}
    \centering
    \includegraphics[width = \linewidth]{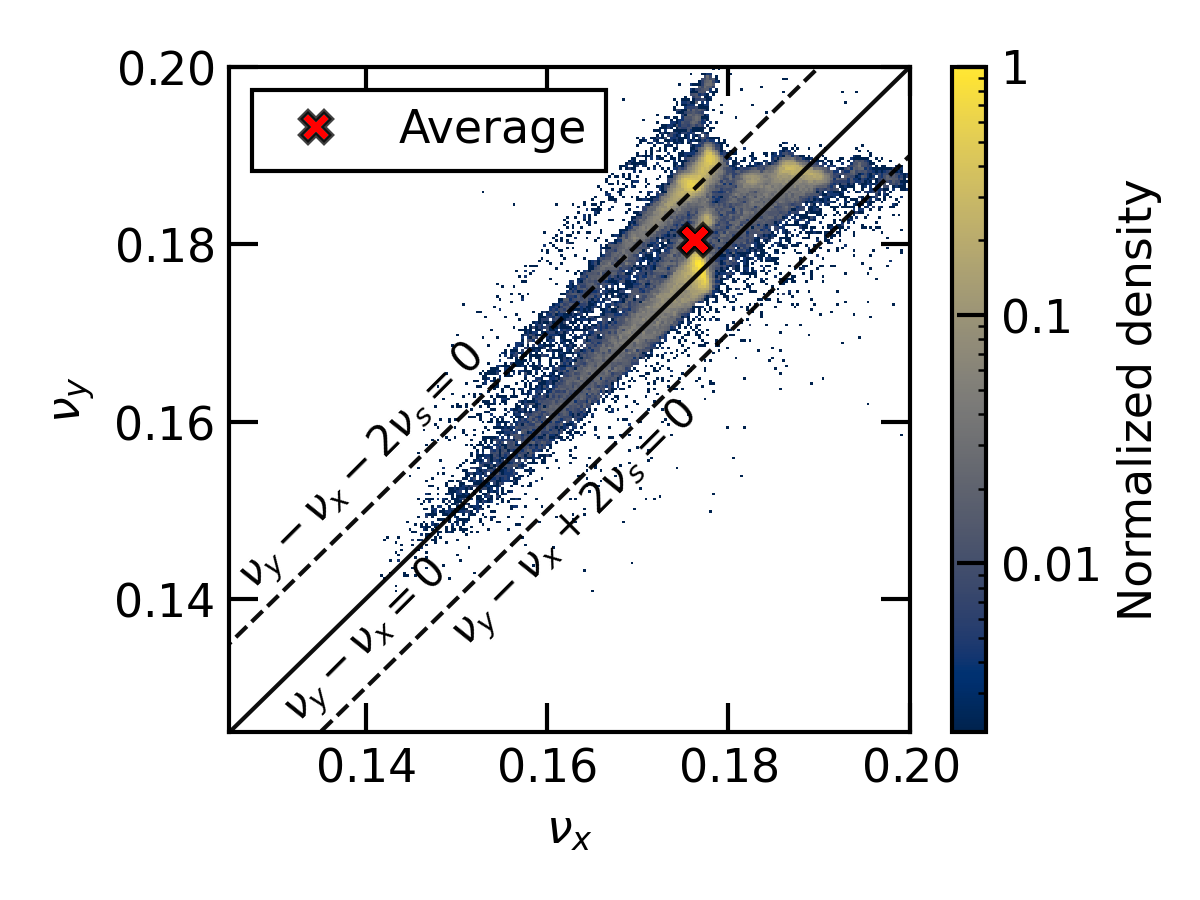}
    \caption{SC tune footprint of a high charge bunch close to the coupling resonance. Bunch charge 10~nC, tune set point: 0.18, 0.185, only the main rf system. Simulation in \texttt{XSuite}~\cite{XSuite}. Particles group themselves around the coupling (solid line) and the SC-driven (dashed lines) resonance. The average incoherent tune is denoted with a cross.}
    \label{fig:tune-footprint-PETRA-IV-coupling}
\end{figure}

\begin{figure}[!h]
    \centering
    \includegraphics[width = \linewidth]{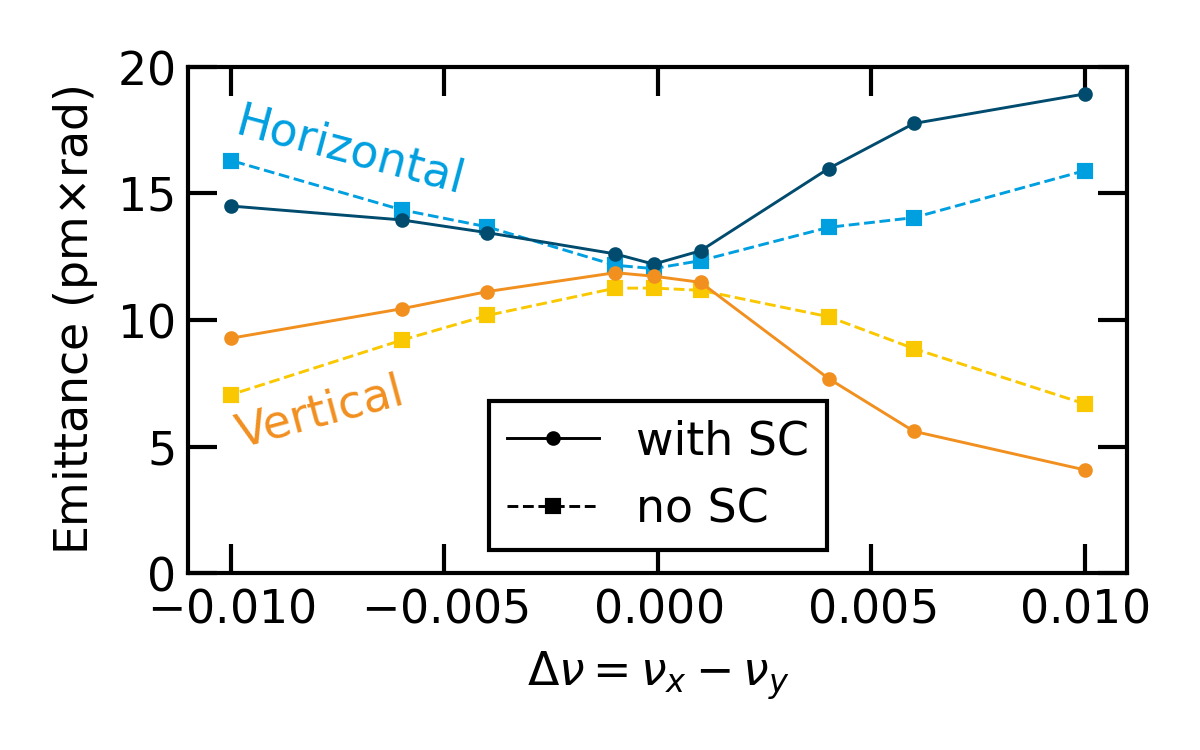}
    \caption{Operating at the coupling resonance could become tricky due to the tune depression. \SI{10}{\nano\coulomb} only main cavity, no wakefields, $\xi = 0$. Lighter dashed lines depict the situation without SC force; darker solid ones -- with SC.}
    \label{fig:SC-CouplingRes}
\end{figure}

Our studies found no harmful effect of the SC-driven resonances on the nominal PETRA~IV working point. At the same time, a significant effect will be present if one decides to operate at a high coupling. Figure~\ref{fig:tune-footprint-PETRA-IV-coupling} shows an exemplary tune footprint for a 10~nC bunch and an emittance coupling ratio close to 1. One can clearly observe the excitation of the SC-driven resonance line, in line with Eq.~(\ref{eq:SC_resonance}). This may result in a change of the emittance ratio, forcing the beam to become more round as it approaches the coupling resonance $\nu_x - \nu_y = 0$ from below~(Fig.~\ref{fig:SC-CouplingRes}). Above, when $\nu_x > \nu_y$, the negative SC tune shift, being larger in the vertical plane, drives the beam away from the coupling resonance, thus making the beam less round than it otherwise would have been. As this change of emittance ratio is charge-dependent, it might complicate the operation of the machine at full coupling. 

\subsection{Collective stability}
Synchrotron light sources rely on a combination of chromaticity, synchrotron radiation damping, and active feedback to ensure single- and coupled-bunch stability. Landau damping is normally negligible due to the small beam size and therefore small amplitude-dependent betatron tune spread within the bunches. Space charge can create a significant amount of betatron tune spread, thus adding Landau damping as an additional tool at disposal of accelerator physicists. In the following Section, we consider the effect of SC on transverse single- and coupled-bunch dynamics.

\subsubsection{Stabilization at chromaticity 0}

\begin{figure}[!h]
    \centering
    \includegraphics[width = 0.99\linewidth]{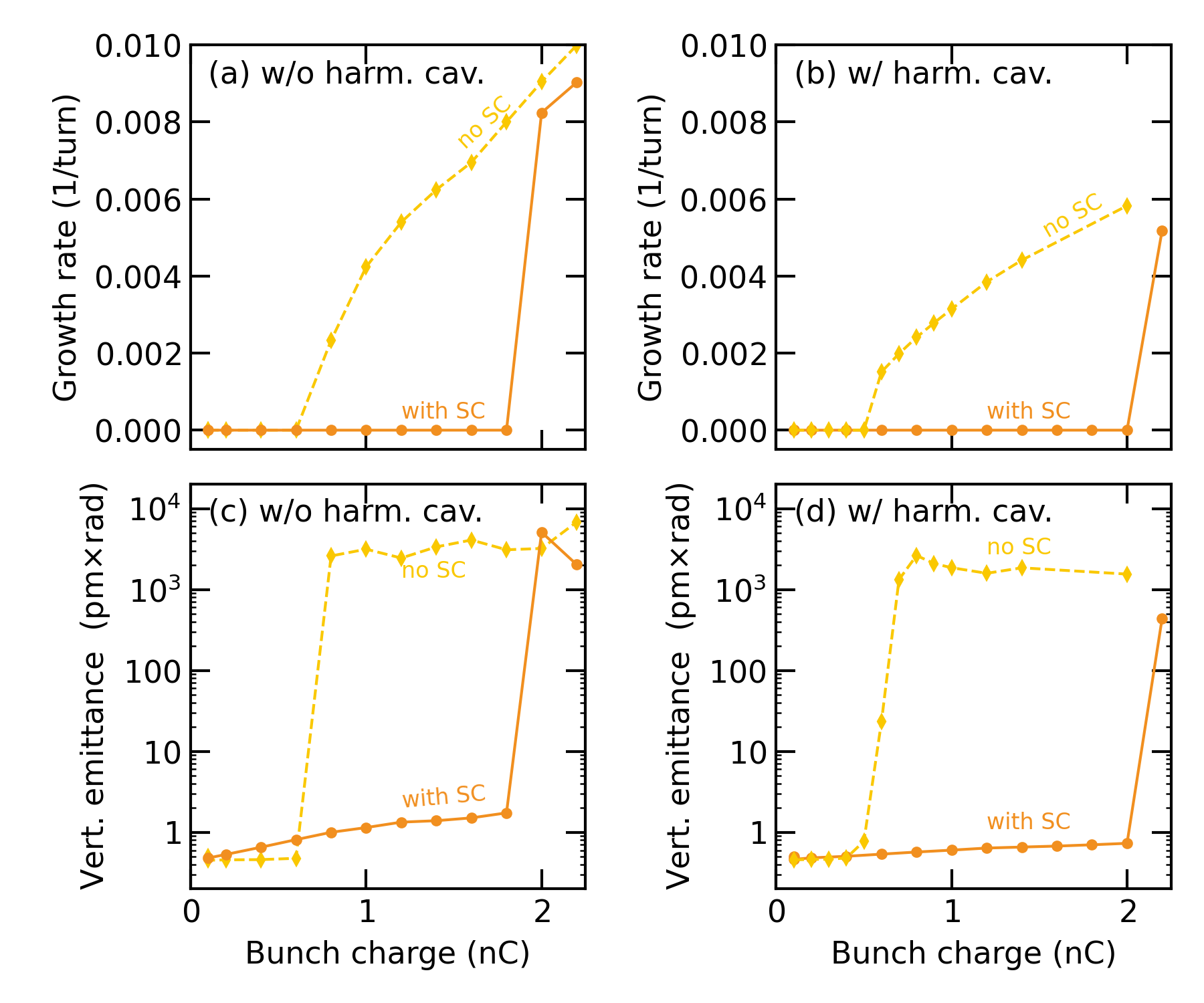}
    \caption{Space charge mitigates TMCI instability at chromaticity 0. The exponential growth rates of the c.\,m. motion in the vertical plane with and without the harmonic rf (a,b) and the final vertical beam emittance after 4096 revolutions, or 20 synchrotron periods, (c,d) as a function of bunch charge at $\xi = 0$. All results were obtained using element-by-element tracking in \texttt{ELEGANT} with a realistic 3\% beta-beating.}
    \label{fig:TMCI_P4}
\end{figure}

In the absence of SC the TMCI threshold at $\xi = 0$ for the PETRA~IV light source ($Z_y \approx 1.3$~M$\Omega$/m) is 0.5~nC without lengthening cavities, according to Eq.~(\ref{eq:TMCI}). Applying Eq.~(\ref{eq:TMCI}) to the case with the 3rd harmonic rf ($\left< \nu_s \right>= 1.7\times 10^{-3}$) we obtain $Q_\text{th} = 0.3$~nC. These limits are in good quantitative agreement with our tracking in \texttt{ELEGANT} (Figs.~\ref{fig:TMCI_P4}a,b), in which we tracked the beam for 4096 turns, or about 1.5 synchrotron radiation damping times, taking into account bunch lengthening by self-induced wakefields.

Including the SC in our tracking simulation we observe that it effectively mitigates the TMCI. Without the 3rd harmonic cavity, neglecting the SC yields a single-bunch charge threshold of about 0.7~nC (Fig.~\ref{fig:TMCI_P4}a) with the bunch becoming unstable in the vertical plane and its beam size rapidly increasing.\footnote{The rapid growth rates, typical to TMCI, were found in tracking by observing the amplitude of the c.\,m. motion with a 200-turn moving window and fitting it with an exponential process.} The final beam emittance is shown in Fig.~\ref{fig:TMCI_P4}c: above the instability threshold the beam emittance is significantly degraded. The SC leads to two effects: first, the instability is stabilized and the intensity threshold increasing to about 1.9~nC, nearly three times higher~(Fig.~\ref{fig:TMCI_P4}a); second, below the new threshold the vertical emittance steadily increases with the bunch charge due to the additional coupling produced by SC~(Fig.~\ref{fig:TMCI_P4}c).

With the bunch lengthening 3rd harmonic rf in place one observes a similar picture. Without the SC force the bunch becomes vertically unstable at about 0.5~nC but when we include the SC it remains stable well above this threshold (Fig.~\ref{fig:TMCI_P4}b). In this case, we do not observe such a significant increase of the vertical emittance with the bunch charge as without the 3rd harmonic rf~(Fig.~\ref{fig:TMCI_P4}d), likely due to the large bunch length and, hence, smaller charge density.

\subsubsection{Single-bunch accumulation limit}
Accumulation of bunch charge during off-axis top-up injection requires that the injection losses, both in the stored and in the injected beams, be relatively small compared to the injected charge. As soon as the injected losses approach the injected charge, further accumulation is not possible.

\begin{figure}
    \centering
    \includegraphics[width = 0.99\linewidth]{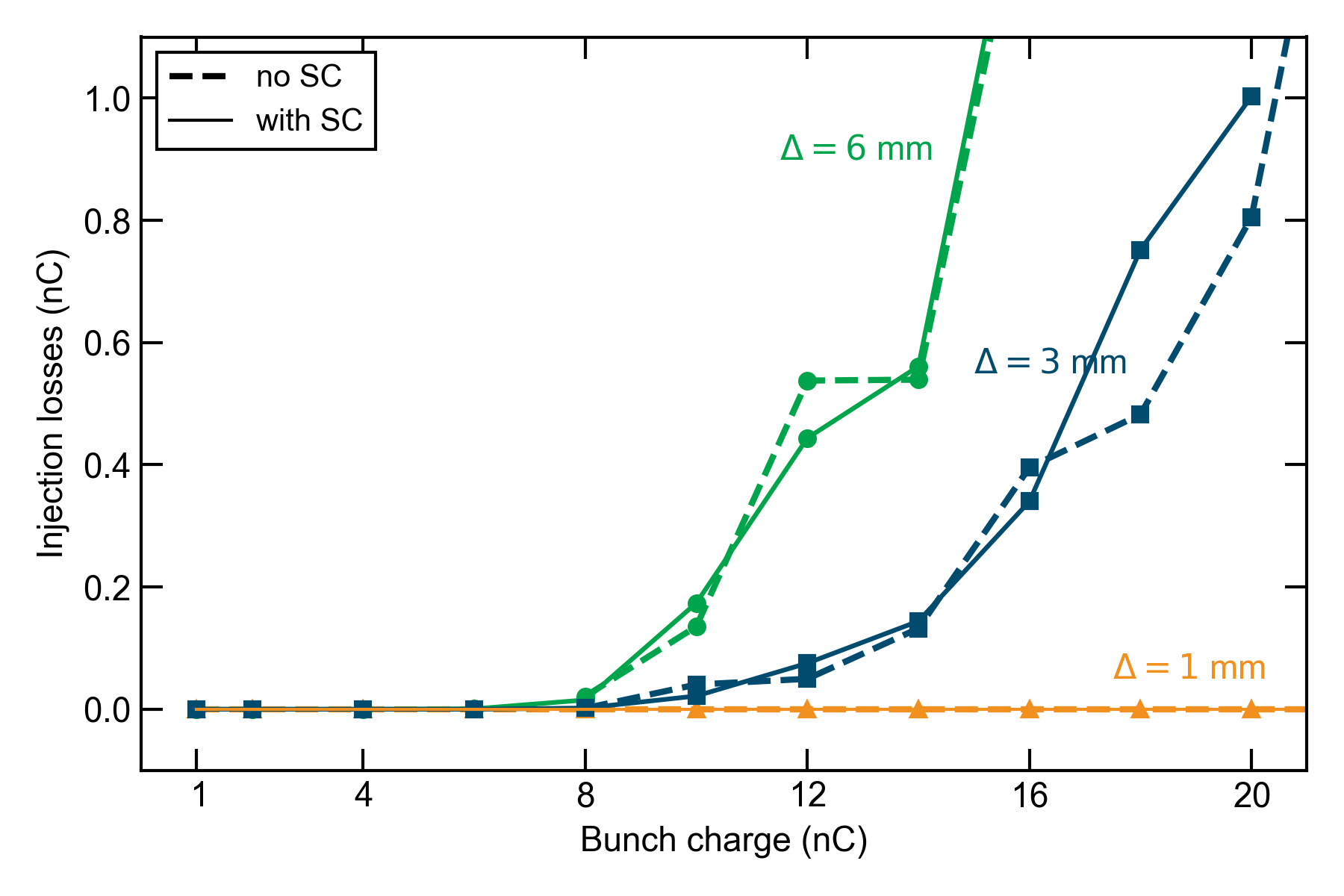}
    \caption{When the stored bunch is significantly displaced from the reference orbit at injection (e.g. during the aperture sharing), the SC has no effect on the resulting losses. Three aperture sharing amplitudes $\Delta$ considered from 1 to 6~mm. Element-by-element tracking in \texttt{ELEGANT} with a realistic aperture model and 3\% beta-beating.}
    \label{fig:Ap_share_small}
\end{figure}

At PETRA IV, significant injection losses happen in the stored bunch when employing aperture sharing, i.e. the injection bump is intentionally not closed to provide additional room for the injected beam. 
After the injection kick the stored beam rapidly decoheres in a few turns and its density drastically decreases. Because of this SC cannot have any significant effect on the injection dynamics when employing aperture sharing. Element-by-element tracking in \texttt{ELEGANT} confirms this hypothesis: no significant difference in the amount of injection losses has been observed when tracking the beam for 4096 turns after the injection~(Fig.~\ref{fig:Ap_share_small}) at $\xi = 5$, which corresponds to an rms head-tail phase advance $\chi = \xi \times 2 \pi \sigma_z / C \alpha_C = 8$.

\begin{figure}[!h]
    \centering
    \includegraphics[width = 0.99\linewidth]{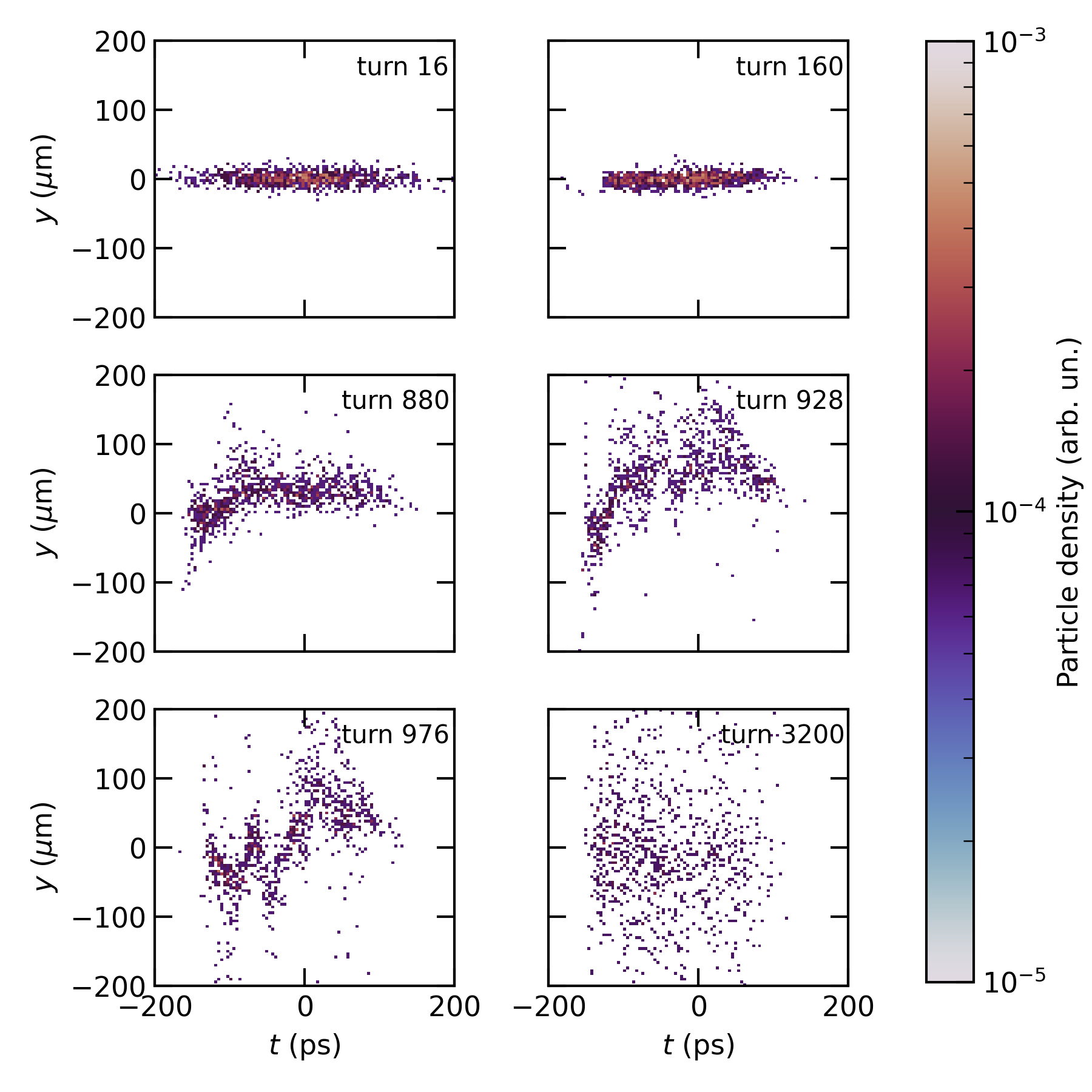}
    \caption{Head-tail instability developing in the vertical plane after injection at turn 0. Bunch charge 14~nC, chromaticity 5, no SC. Color denotes the beam density on a logarithmic scale.}
    \label{fig:HT_Instab}
\end{figure}

\begin{figure}[!h]
    \centering
    \includegraphics[width = 0.99\linewidth]{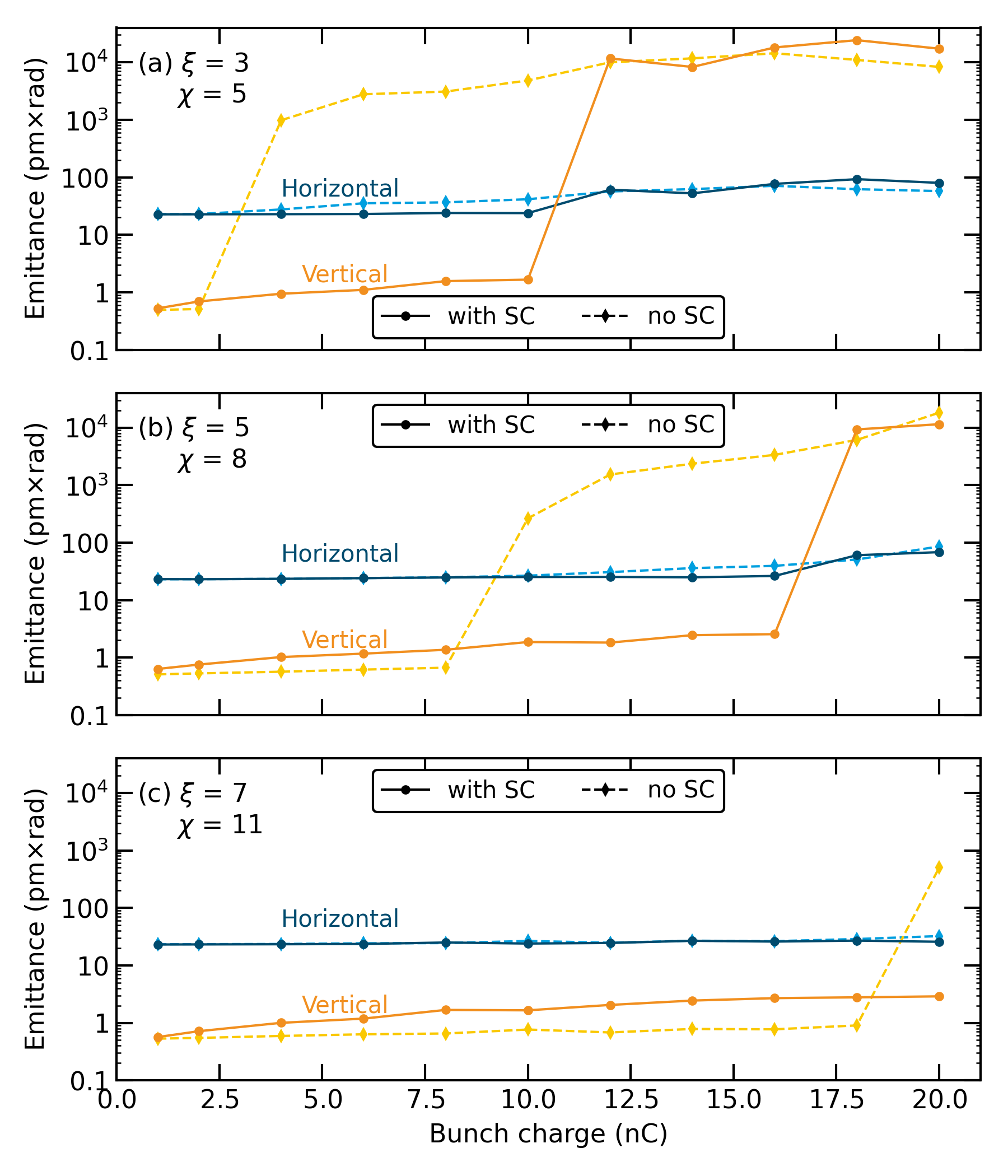}
    \caption{As the bunch charge increases the vertical emittance is blown up by the head-tail instability. Space charge helps stabilizing the beam, increasing the single bunch current limit. Chromaticity 3 (top), 5 (middle), and 7 (bottom) with 3rd harmonic rf and 3\% beta-beating.}
    \label{fig:Acc-limits}
\end{figure}

At the same time, even if the losses are not considerable the beam quality might be compromised by a head-tail instability triggered by the injection event~(Fig.~\ref{fig:HT_Instab}). The instability rapidly increases the beam size to about 100 times the initial one. Further blow-up and particle losses do not happen, which can be attributed to the nonlinear amplitude detuning providing sufficient Landau damping to suppress the instability growth. Such a blown-up beam is doubtfully useful for the photon science because of its low brightness. The SC provides an additional stabilization mechanism and helps mitigating the head-tail instability (Fig.~\ref{fig:Acc-limits}). This allows accumulating a greater bunch charge without compromising the beam quality.

Thus, in the presence of space charge our dynamical system may have two distinct equilibrium states: the low-emittance state, where it is stabilized by the nonlinearity of the space charge, and the high-emittance state, where it is stabilized by the nonlinear amplitude detuning. The former state is unstable in nature, so a small increase of beam emittance can lower the stabilizing space charge force, reducing the strength of Landau damping and driving the beam unstable, further increasing the emittance. The later is a stable equilibrium state as a small increase of beam emittance would only further increase the nonlinear tune spread, enhancing the Landau damping.

\begin{figure}[!h]
    \centering
    \includegraphics[width = 0.99\linewidth]{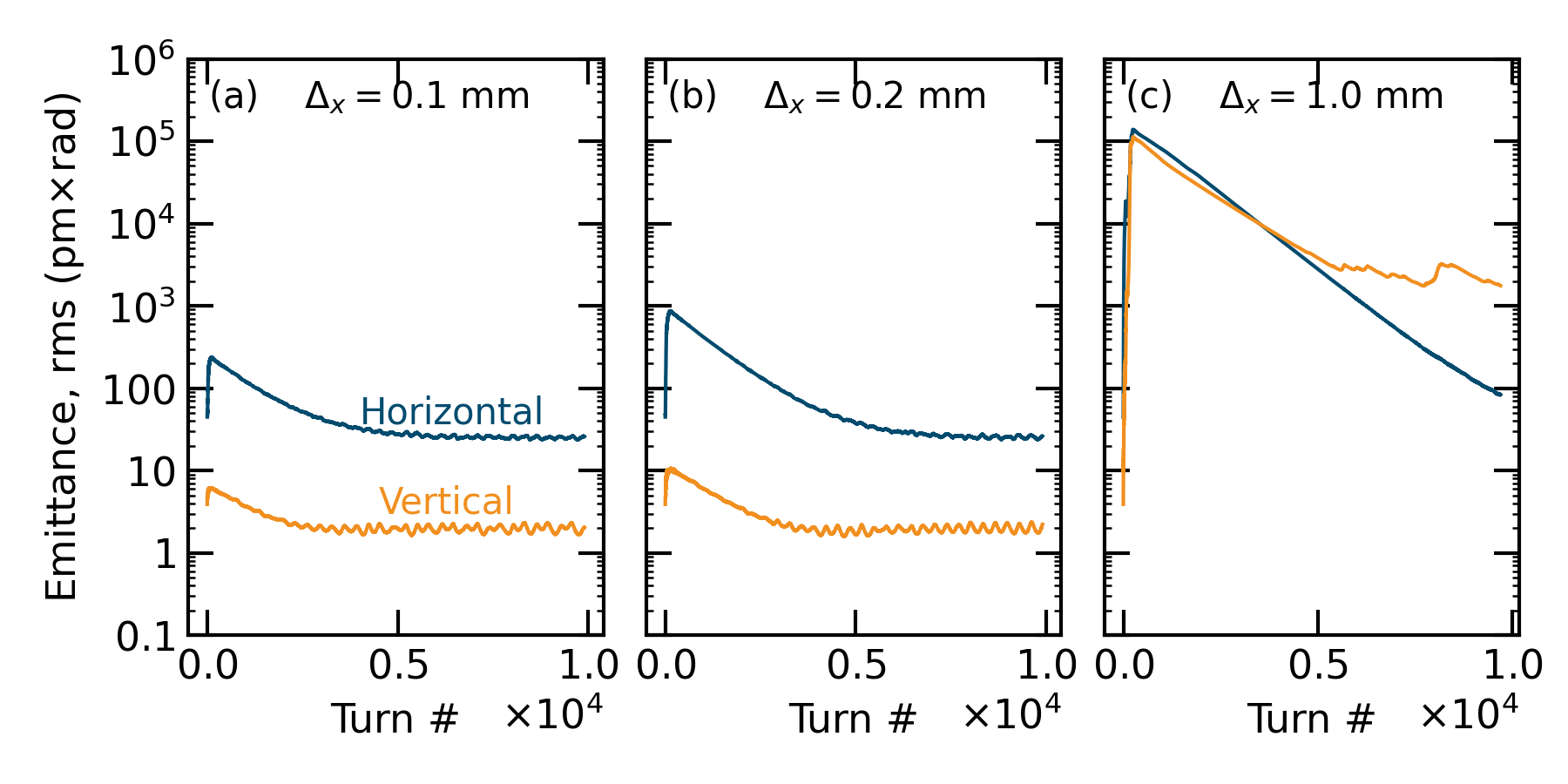}
    \caption{Evolution of the rms emittance of the stored bunch after a small parasitic kick during the top-up injection (a, b). The injection kick of a large enough amplitude might destabilize the bunch that has been kept stable by SC, blowing up its emittance (c). Bunch charge 12~nC, chromaticity 5, 3\% rms beta-beating.}
    \label{fig:inj_perturb}
\end{figure}

To illustrate that the high-bunch-charge state stabilized by SC is an unstable equilibrium let us consider small injection perturbations of the stored bunch that might result from an imperfect closure of the injection bump. At PETRA IV the injection bump amplitude is 10~mm, and it is estimated that the field quality in the kickers as well as the pulse shape stability in the power supplies guarantee a relative stability of the injection kick at the $10^{-3}$ level~\cite{loisch:priv_comm_2022}. That is, we do not expect injection bump non-closures far beyond 100~$\mu$m in operation. Yet, for the sake of making an argument we will consider larger values here. Let us consider a bunch that should be unstable without space charge but is stabilized by it, e.g. 12~nC at $\xi = 5$ (Fig.~\ref{fig:Acc-limits}b). Assume there is an injection bump non-closure in the $x$-plane. Due to non-zero coupling between the transverse planes the bunch will execute small oscillations in both planes after injection, resulting in an increase of its projected emittances. For small bump amplitudes, ca. $100-200~\mu$m the bunch remains stable and the synchrotron radiation damps the emittances down to their unperturbed equilibrium values (Fig.~\ref{fig:inj_perturb}a,b). When the non-closure is large, i.e. 1~mm (Fig.~\ref{fig:inj_perturb}c), the dynamics is altered. The SC no longer stabilizes the bunch and the vertical emittance settles at the value it would have if there were no SC, about $10^4$~pm$\times$rad, three orders of magnitude larger than that of an unperturbed bunch.  

In order to prevent the degradation of beam quality one may consider increasing the chromaticity, which keeps the bunch stable without the SC (Fig.~\ref{fig:Acc-limits}c). But for the sake of a larger DA and a longer beam lifetime it might be beneficial to operate at lower $\xi$. Then one may take advantage of the stabilization offered by the SC by keeping the chromaticity at an optimal, potentially lower setting during the fill and increasing it only temporarily during top-up injections.




\section{SOLEIL~II case}\label{sec:S2_case}
Since SOLEIL~II plans using MIK for injection~\cite{alex:ipac23-thpa175}, this avoids the need for aperture sharing, and therefore, no collective effects are expected at injection. Yet, the SC might still have an effect on the stored beam. 
The expected space-charge tune shifts in SOLEIL~II are relatively large and are displayed in Table~\ref{tab:SC_tune_shifts}.

\subsection{Simulation setup}
The primary tool for studying collective effects at SOLEIL is particle tracking code \texttt{mbtrack2}\,\cite{gamelin_mbtrack2_2024, gamelin_mbtrack2_2021}. 
The codes allow us to simultaneously consider single- and coupled-bunch collective effects in transverse and longitudinal planes. 
A linearized Vlasov equation solver \texttt{DELPHI}\,\cite{mounet_landau_2020} is used additionally if required.
Transverse space charge was recently implemented in \texttt{mbtrack2} using a Basetti-Erskine formula \cite{bassetti_closed_1980} with a so-called matched frozen space charge model~\cite{kornilov_beam_2020}. 
The longitudinal bunch profile is used to modulate the strength of the space charge kick. 

Simulations in this paper are based on the lattice parameters reported in \cite{loulergue_tdr_2023}.
Simulations in \texttt{mbtrack2} are dedicated to collective effects, so a few simplifications are made. 
All simulations include the following elements: a one-turn map for transverse motion, a detuning model of chromaticity, synchrotron radiation effects, transverse short and long-range wakefields, longitudinal short-range wakefield, rf cavities setup and a lumped space-charge kick. 
We use the full impedance model of SOLEIL~II with more than \num{50} distinct elements\footnote{The current impedance model of SOLEIL~II does not yet include quadrupolar wakefields.
As the vacuum chamber of SOLEIL~II is round, many of the impedance model components are axisymmetric, except for the in-vacuum IDs; it is expected that quadrupolar wakes are less important in SOLEIL~II than they are in 3rd-generation light sources.}. 
The geometric part of the impedance was computed with \texttt{CST Particle Studio} and \texttt{ABCI}. 
NEG-coated resistive wall and image charge contributions were obtained with \texttt{IW2D}. 
In all cases, for the first \num{25000} turns, only longitudinal wakefield acts on the beam to reach an equilibrium bunch length. 
After which transverse wakes are turned on, and the beam is tracked for another \num{25000} -- \num{75000} turns. 
\num{1000000} macroparticles per bunch are used in all simulations. 
The bunch is partitioned into \num{100} longitudinal slices for wakefield and space-charge interactions.

Results without space charge, the corresponding simulation setup, the impedance model of SOLEIL~II, the current harmonic cavity configuration, and transverse feedback requirements were discussed in detail in \cite{gubaidulin_transverse_2024}. 
The effect of IBS is only included when specified.
IBS is modelled in tracking using a simple kick model~\cite{bruce_time_2010} based on an analytical calculation of the IBS growth rate at every turn.
We have employed the completely integrated Piwinski model~\cite{kubo_intrabeam_2005} for growth rate calculation.
This paper considers only vertical instabilities for SOLEIL~II with closed gaps of insertion devices to obtain the strongest instabilities.
The growth rates of all considered instabilities are found by fitting the beam c.\,m. offset with an exponential function.

\subsection{Single bunch instabilities}

\subsubsection{Transverse mode-coupling instability}
\begin{figure}
    \centering
    \includegraphics[width=\columnwidth]{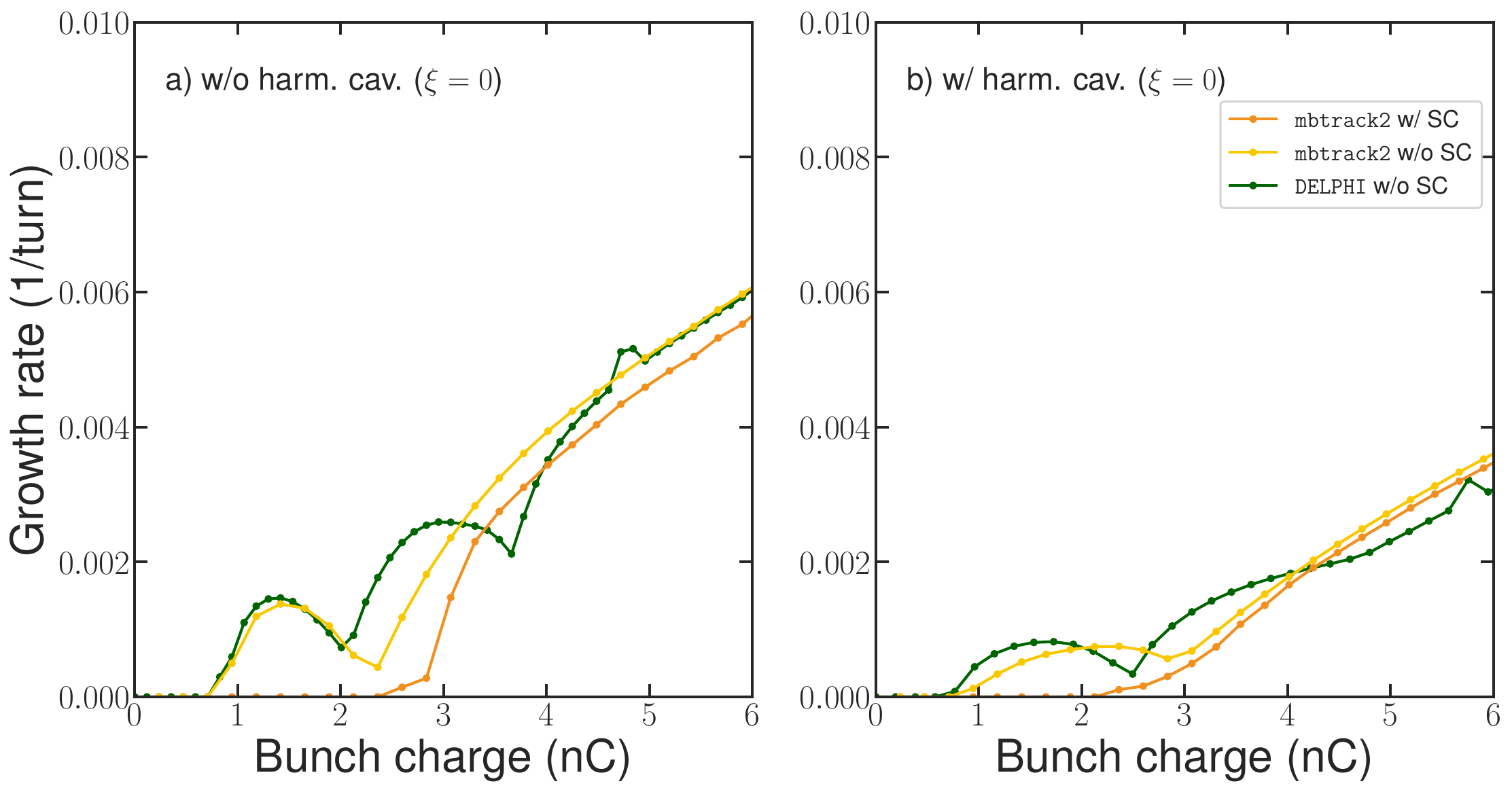}
    \caption{
    Mitigation of TMCI by space-charge for the case of SOLEIL~II.
    The instability growth rates are plotted against the bunch current. 
    a) the case without a harmonic cavity; b) the case with a harmonic cavity.}
    \label{fig:soleil_tmci}
\end{figure}

Coupling of the azimuthal modes \num{-1} and \num{0} causes a transverse mode coupling instability.
Figure~\ref{fig:soleil_tmci} compares growth rates of TMCI with and without space charge for two cases. 
One case (Fig.~\ref{fig:soleil_tmci}a) shows a configuration without a harmonic cavity, and another (Fig.~\ref{fig:soleil_tmci}b) shows one with one included. 
In both cases, space-charge significantly increases the threshold current of TMCI by more than twofold. 
It is well known \cite{ABS_Model} that space-charge can delay or even prevent the coupling of azimuthal modes \num{-1} and \num{0}.

\begin{figure}
    \centering
    \includegraphics[width=\columnwidth]{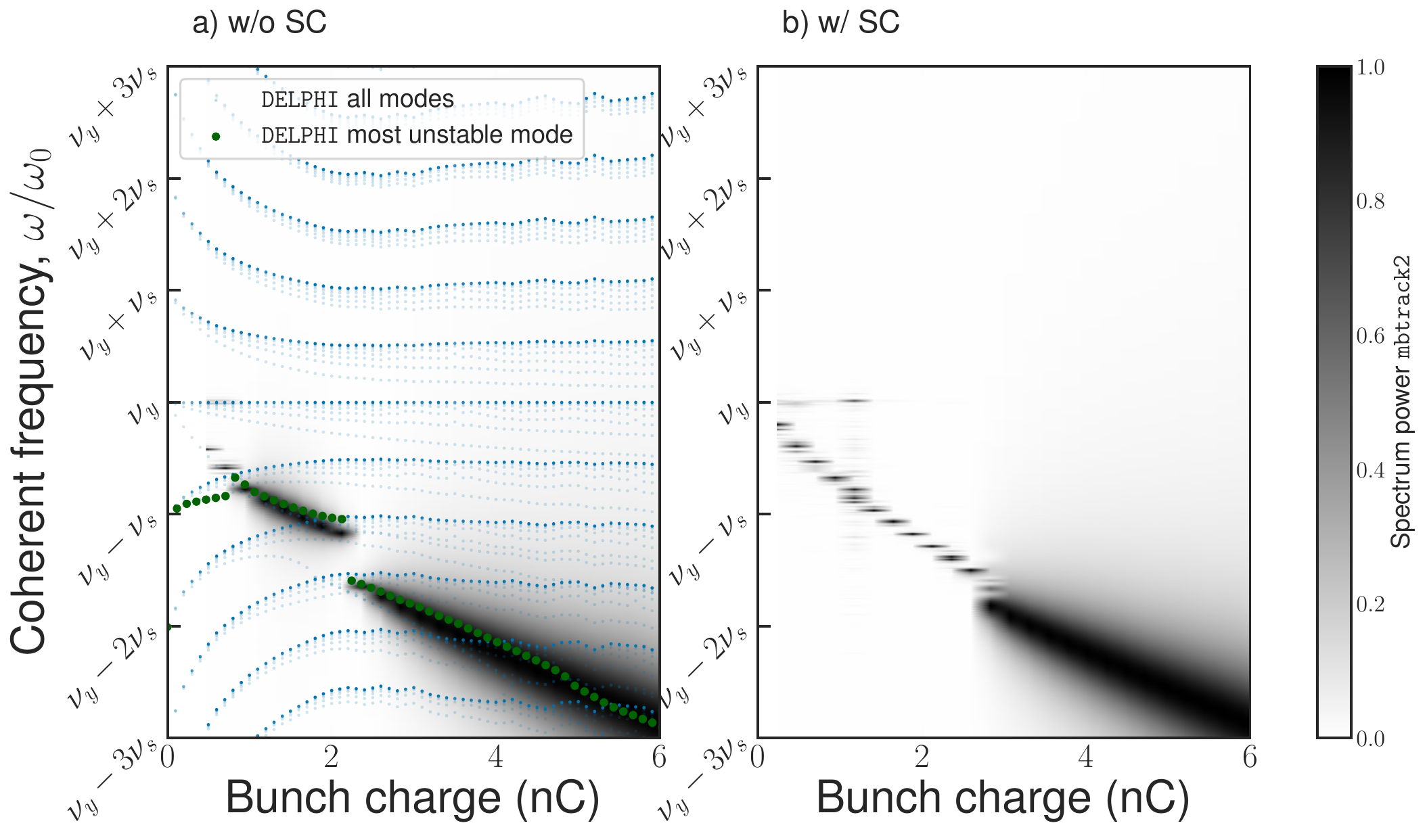}
    \caption{Coherent frequencies of oscillations before and after the TMCI threshold w/o harmonic cavity.
    a) Case w/o SC, compared to analytical solutions of \texttt{DELPHI}.
    b) Case w/ SC.
    }
    \label{fig:soleil_tmci_real_part}
\end{figure}
Figure~\ref{fig:soleil_tmci_real_part} demonstrates the coherent frequencies of bunch oscillations as a function of bunch charge. 
In the case without space-charge (Fig.~\ref{fig:soleil_tmci_real_part}a), modes \num{-1} and \num{0} couple at $\sim\SI{0.9}{\nano\coulomb}$ and determine the TMCI threshold. 
At an even higher charge $\sim\SI{2}{\nano\coulomb}$ modes \num{-3} and \num{-2} couple and drive the instability.

With space charge (Fig.~\ref{fig:soleil_tmci_real_part}b), the TMCI threshold is now at $\sim\SI{2}{\nano\coulomb}$. 
After the threshold, the bunch oscillates at similar frequencies as in the case without space charge for the same bunch charge.
This can be interpreted as follows. 
Space charge prevented coupling of modes \num{-1} and \num{0}. 
But the modes \num{-3} and \num{-2} still couple at a similar bunch charge regardless of space charge.

\subsubsection{Head-tail instability}
\begin{figure}
    \centering
    \includegraphics[width=\columnwidth]{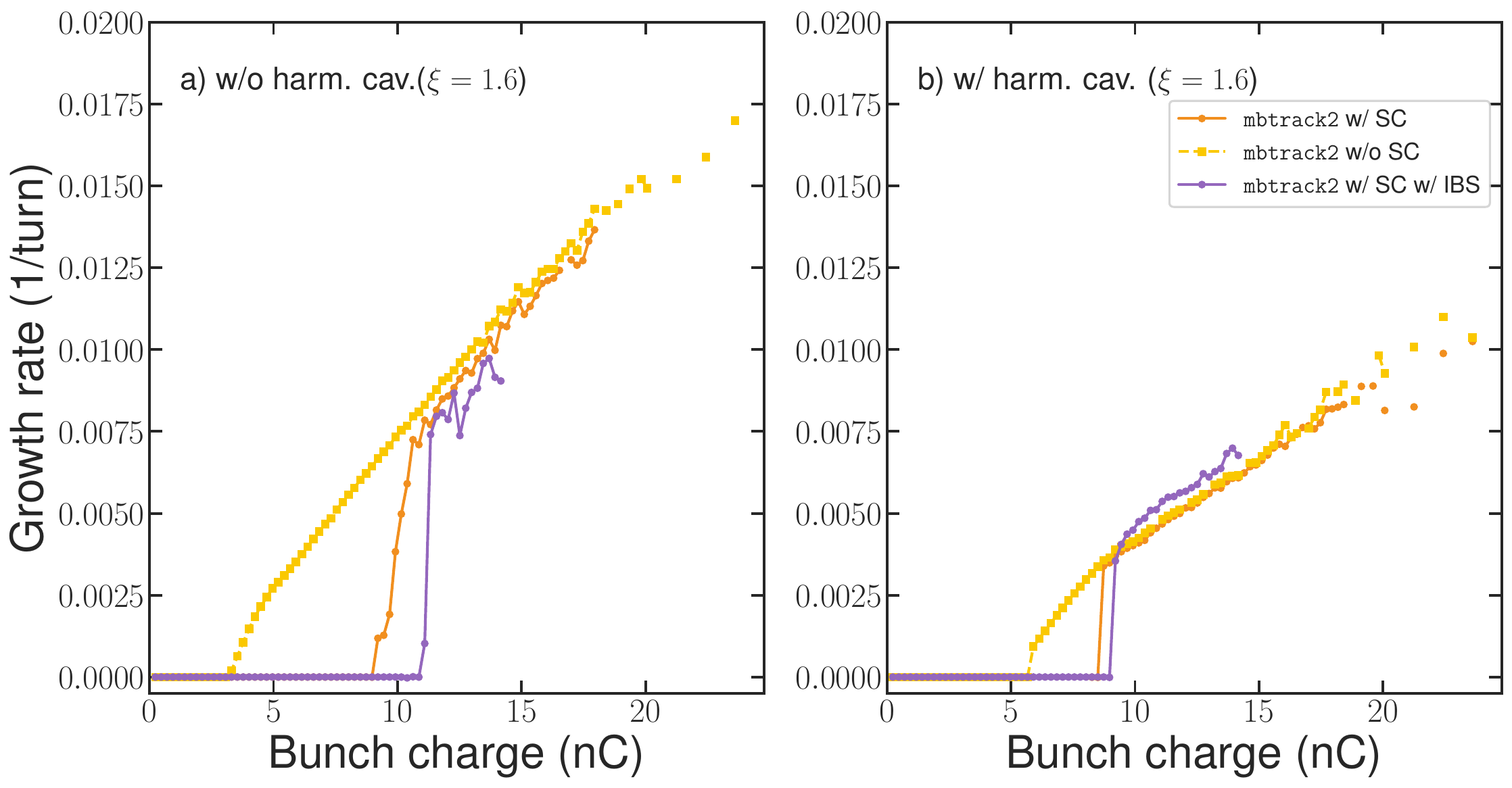}
    \caption{
    Suppression of head-tail instability for the case of SOLEIL~II.
    The instability growth rate is plotted against the bunch current.
    a) the case without a harmonic cavity; b) the case with a harmonic cavity.}
    \label{fig:soleil_head-tail}
\end{figure}

Figure~\ref{fig:soleil_head-tail} compares growth rates of head-tail instability with and without space charge.
Two cases are again considered at the $\xi = 1.6$ design chromaticity.
One without including a harmonic cavity ($\chi=1.1$ at $Q=\SI{1.4}{\nano\coulomb}$) and the other with a harmonic cavity ($\chi=3.3$ at $Q=\SI{1.4}{\nano\coulomb}$). 
In both cases, space charge completely suppresses head-tail instability up to $\sim\SI{8}{\nano\coulomb}$.
The instabilities are suppressed by Landau damping due to the space charge of a bunched beam. 
This strong suppression of head-tail instability by space-charge (for the intermediate space-charge parameter values $|\Delta\nu^\text{SC}_y|/\nu_s\in [2, 20]$) is qualitatively consistent with results obtained for hadron synchrotrons analytically \cite{balbekov_transverse_2009, burov_head-tail_2009, burov_erratum_2009} and in simulations \cite{kornilov_head-tail_2010, macridin_simulation_2015}.

Figure~\ref{fig:soleil_head-tail} also compares growth rates obtained with space charge for two scenarios: with and without including IBS in the simulation.
For Fig.~\ref{fig:soleil_head-tail}a, there is a notable difference in obtained threshold ($\sim\SI{8}{\nano\coulomb}$ without IBS and $\sim\SI{11}{\nano\coulomb}$ with IBS). 
This can be interpreted in the following manner.
From hadron machine research, it is known that space charge provides Landau damping in bunches only for the intermediate space charge parameter. 
In our case, we assume that the instability threshold is determined by the loss of Landau damping due to the high value of the space-charge parameter.
IBS will increase the beam emittance, thus reducing space-charge tune shift and increasing the instability threshold.
For Fig.~\ref{fig:soleil_head-tail}b, a similar but smaller increase of the threshold bunch charge is observed. 
This corresponds to the weaker IBS effect for bunches lengthened by a harmonic cavity.


\subsection{Transverse coupled-bunch instability}

\begin{figure}
    \centering
    \includegraphics[width=\columnwidth]{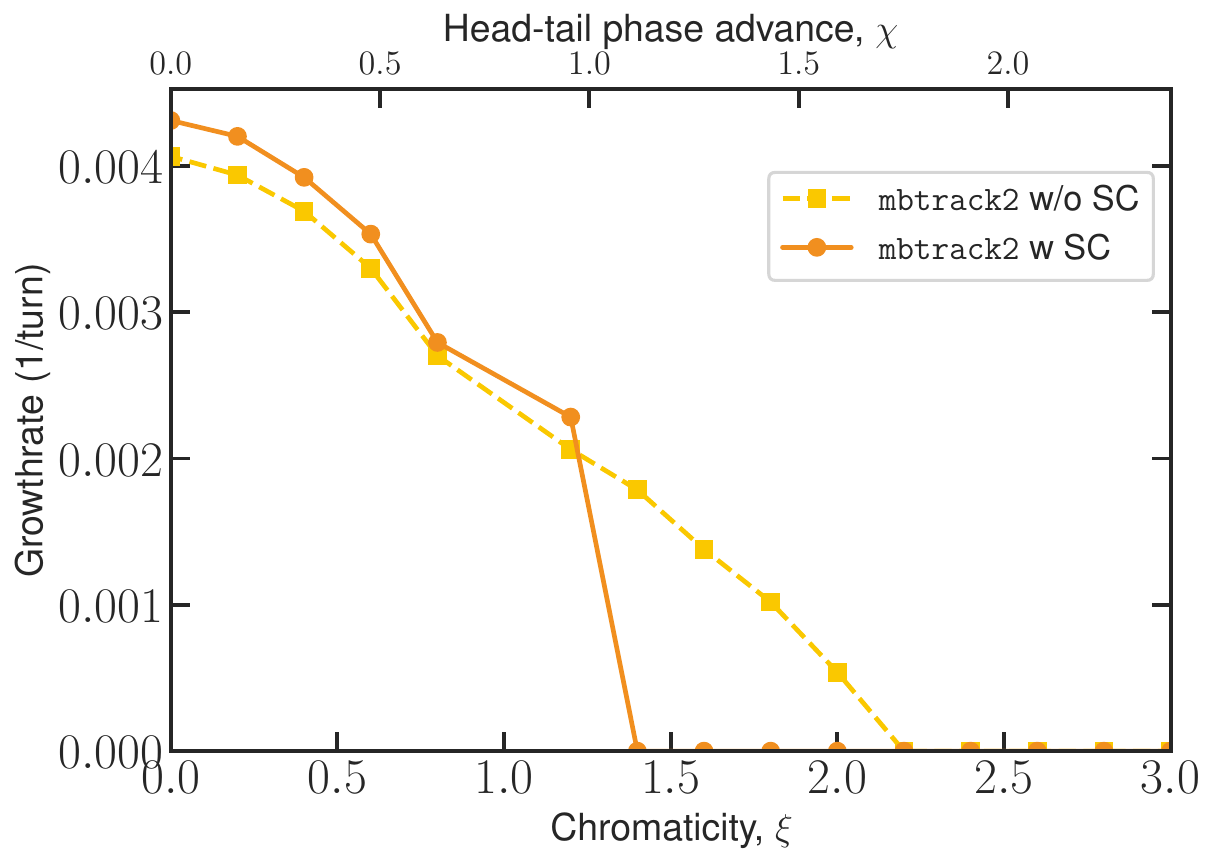}
    \caption{Mitigation of TCBI in SOLEIL~II by a combination of space-charge and chromaticity w/o harmonic cavity.}
    \label{fig:soleil_tcbi}
\end{figure}

Figure~\ref{fig:soleil_tcbi} shows the growth rates of transverse coupled-bunch instability (TCBI) for SOLEIL~II Brightness operation mode with \num{416} bunches at \SI{500}{\milli\ampere} for different setting of chromaticity $\xi$ for the case without a harmonic cavity.
It demonstrates that TCBI is suppressed for chromaticities $\xi > 1.5$ due to space charge. 
At these higher chromaticities, transverse coupled-bunch instability is already determined by both intrabunch and interbunch motions \cite{skripka_simultaneous_2016}. 
Space charge does not affect pure TCBI at $\xi = 0$ with $m=0$ intrabunch mode, as seen from very close growth rates in Fig.~\ref{fig:soleil_tcbi}. 
However, at higher chromaticities, this mode is known \cite{cullinan_transverse_2016} to be overtaken by higher modes ($ m\geq1$) that can be affected by space charge.  

\begin{figure}
    \centering
    \includegraphics[width=\columnwidth]{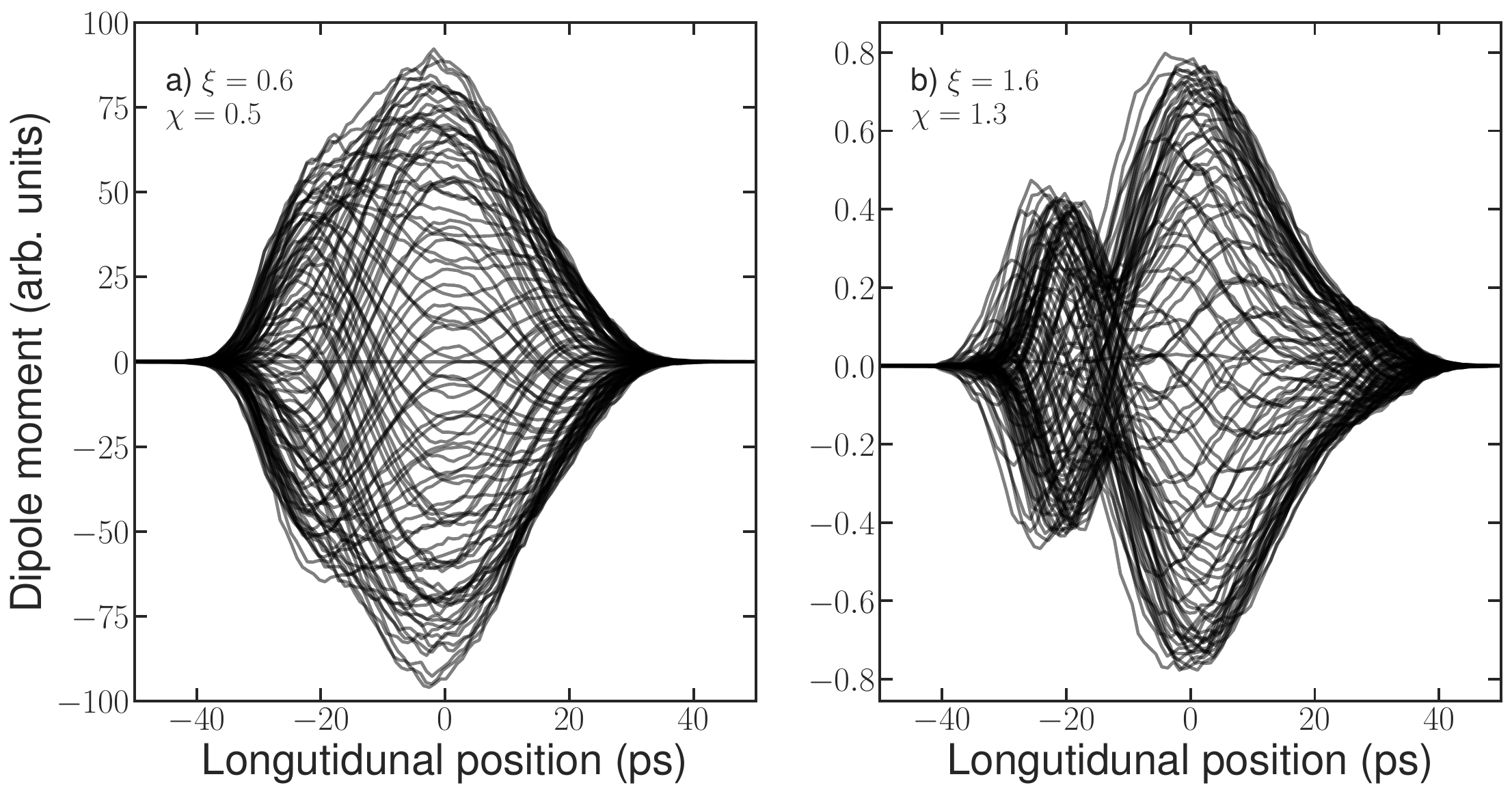}
    \caption{Intrabunch motion (dipolar moment) of one of the bunches for a) $\xi=0.6$ and b) $\xi=1.6$ w/o SC. 
    Obtained from \texttt{mbtrack2} simulations of Fig.~\ref{fig:soleil_tcbi} for \SI{1.4}{\nano\coulomb} bunch charge.
    }
    \label{fig:soleil_intrabunch_tcbi}
\end{figure}

Indeed, Fig.~\ref{fig:soleil_intrabunch_tcbi} demonstrates two different intrabunch modes $|m|=0$ (at $\xi=0.6$) and $|m|=1$ (at $\xi=1.6$) for the case without space-charge. 
We considered the Brightness mode of SOLEIL~II with \num{416} equidistant bunches at \SI{1.4}{\nano\coulomb} equivalent to \SI{500}{\milli\ampere} beam current.
For the Timing mode of SOLEIL~II, the coupled-bunch instability is already suppressed by synchrotron radiation.
As expected at low chromaticity (e.g. $\xi=0.6$), intrabunch motion corresponds to $|m|=0$ mode. 
At higher chromaticity (e.g. $\xi=1.6$), it corresponds to $|m|=1$ mode instead.
At low chromaticities, the growth rates of the instability with and without space charge are very similar. 
When instability is dominated by $|m|=1$ mode, space-charge suppresses it.
This demonstrates that space charge can affect coupled-bunch instabilities for sufficiently high chromaticities.
Therefore, it can help to relax the requirements for a transverse feedback system that is usually employed to suppress this instability.

\section{Conclusion}
In this paper, we have studied the role of space charge in future 4th generation light sources. Thanks to their small transverse beam size, these light sources attain significant space charge tune shifts with the space charge parameter approaching values as high as $\sim 10$ units. This results in a wide variety of new physics that is not typically observed in contemporary electron machines.

On the positive side, SC mitigates the mode coupling instability of modes 0 and 1 and increases the single bunch current threshold at chromaticity 0, as seen in numerical simulations of PETRA~IV and SOLEIL~II machines. At higher positive chromaticities, SC increases the single bunch intensity limits. It can also stabilize the coupled bunch motion, lowering the required chromaticity setting as observed for the \num{416}-bunch mode of SOLEIL~II.
This additional instability mitigation by space-charge can reduce the requirements for a transverse feedback system.
The beam stabilization with space charge can be affected by IBS, especially when operating without a harmonic rf system. Depending on beam parameters, the IBS effect might be beneficial for further increasing the stability threshold, as observed in SOLEIL~II. 
A harmonic rf system, normally designed to reduce the effect of IBS as much as possible, greatly diminishes its impact on beam instabilities.

On the other hand, SC also generates transverse coupling leading to rounder beams at high intensities. As we observe in our PETRA~IV simulations, the large SC tune shift might interfere with the dynamics of the top-up injection. SC can also excite particular resonance lines, increasing the transverse emittance. This might compromise the beam quality, e.g. when operating close to a coupling resonance. Finally, the large SC tune shift might complicate the modes of machine operation with high single-bunch charges when not employing bunch lengthening cavities as seen with the Timing mode of PETRA~IV.

The fact that the stabilizing effect of SC is observed both in medium and high energy light sources and in different simulation setups, as well as expected from the theory, dictates that SC has to be taken into account if one wishes to accurately describe beam dynamics in future low emittance generation light sources.

\section*{Acknowledgements}
SA and VG thank the organizers of the I.FAST~LER'24 workshop for providing an excellent venue for sharing ideas and many fruitful discussions, which ultimately led to writing of this paper.
SA expresses his gratitude to Chao Li for providing the equilibrium PETRA~IV beam parameters with IBS and to Yong-Chul Chae for numerous useful suggestions. This research was supported in part through the Maxwell computational resources operated at Deutsches Elektronen-Synchrotron DESY, Hamburg, Germany.
Part of this work was performed using CCRT HPC resource (TOPAZE supercomputer) hosted at Bruyères-le-Châtel, France.

\appendix





\section{Convergence test}\label{App:B}

\begin{figure}
    \centering
    \includegraphics[width=\columnwidth]{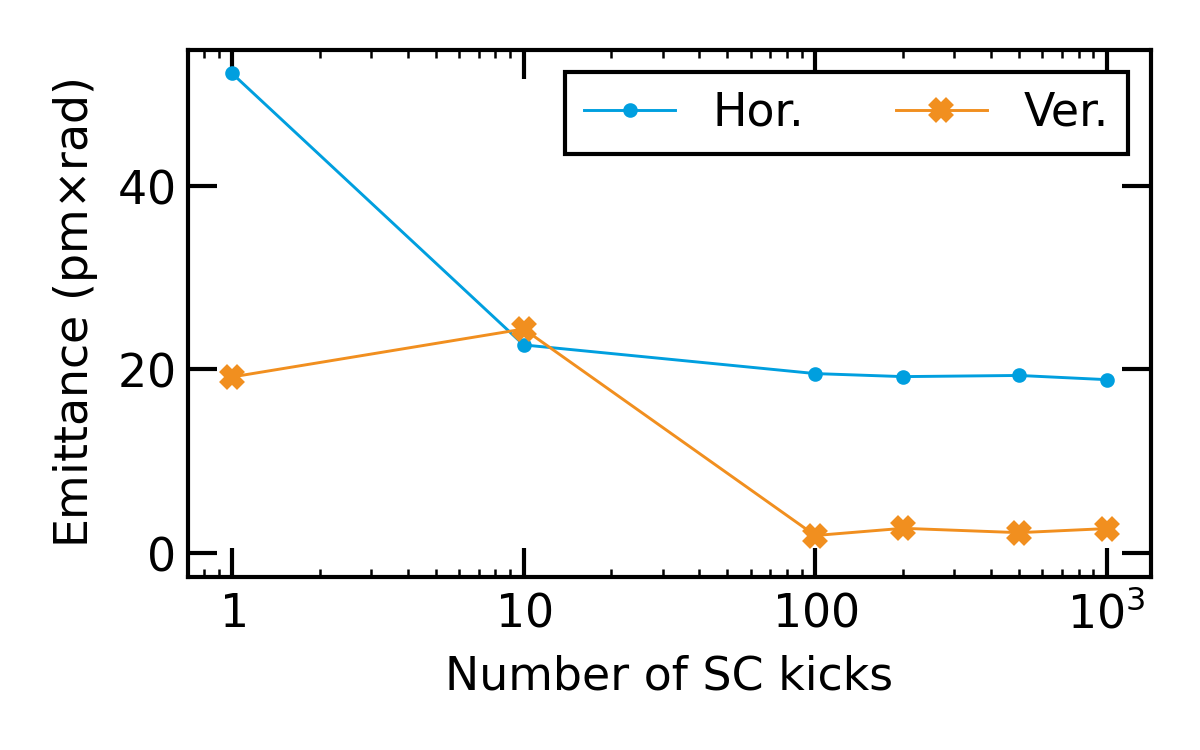}
    \caption{Emittance after 10$^4$ turns of tracking with different amount of equidistantly placed SC kicks. Bunch charge 10~nC, no transverse wakefields.}
    \label{fig:NkicksConvTest}
\end{figure}

The minimum number of SC kicks required for convergence was assessed by introducing an increasing amount of SC kicks equidistantly in the PETRA~IV lattice. The results are shown in Fig.~\ref{fig:NkicksConvTest}. After introducing one SC kick every 23~m, or 1/100th of the machine circumference, the final beam emittances converge to stable values and do not change with the addition of extra SC elements. In our numerical scans in \texttt{ELEGANT} we used about 700 SC kicks with the average distance between them being about 4~m.

\section{Benchmarks with a particle-in-cell code \texttt{XSuite}}
\label{App:C}

\begin{figure}[h!]
    \centering
    \includegraphics[width=\columnwidth]{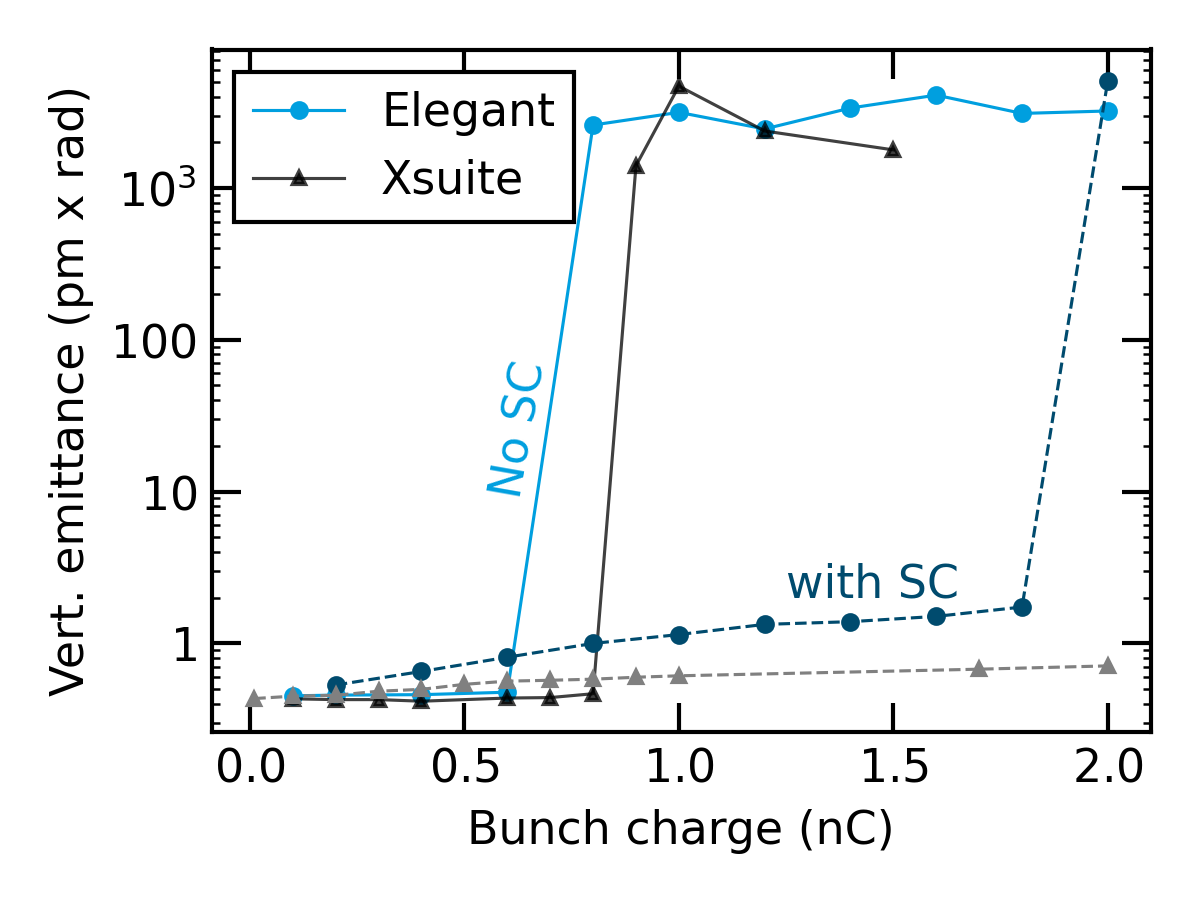}
    \caption{Charge scan comparison between \texttt{ELEGANT} (dots) and \texttt{XSuite} (triangles). Only main rf system is included, $\xi = 0$. Results without (solid curve) SC and with (dashed) SC kicks are shown. }
    \label{fig:ElegantXsuiteComparison}
\end{figure}

In order to benchmark our quasi-frozen SC model in \texttt{ELEGANT} we performed a benchmark comparison with a particle-in-cell (PIC) model, implemented in \texttt{XSuite}. The PIC SC kicks were introduced equidistantly along the lattice every 46m (50 SC kicks in total).
The typical setup for the PIC solver was: (a) a transversal grid divided into $256 \times 256$ segments, range between $\pm$10 rms beam sizes; (b) 105~\unit{\milli \meter}-long longitudinal grid with 100 divisions. 
Details on the implementation of the 2.5D PIC solver can be found in \cite{OeftigerPhDThesis}.
Particles were tracked for 10$^4$ turns, including the effects of synchrotron radiation, quantum excitation, and wakefields. The chromaticity was fully corrected and only the main cavity was included.
The results of this benchmark are illustrated in Fig.~\ref{fig:ElegantXsuiteComparison}.
The data shown in Fig.~\ref{fig:TMCI_P4} (c) are included as blue dots. 
The charge scan simulations are consistent with the results presented in Sect.~\ref{sec:P4_case}. Stabilization of TMCI is confirmed in both numerical models.

\bibliography{biblio} 

\end{document}